\documentclass[a4paper,11pt]{article}
\usepackage{jheppub}
\usepackage[T1]{fontenc}
\usepackage[utf8]{inputenc}
\usepackage{units}
\usepackage{verbatim}
\usepackage{url}
\usepackage{amsthm}

\preprint{MS-TP-23-44}

\title{Anomaly-free dark matter models with one-loop neutrino masses and a gauged U(1) symmetry}

\author[a]{T. de Boer,}
\author[b]{M. Klasen,}
\author[b]{and S. Zeinstra\footnote{Now with Demcon Multiphysics.}}

\affiliation[a]{Max-Planck-Institut f\"ur Kernphysik, Saupfercheckweg 1, 69117 Heidelberg, Germany}
\affiliation[b]{Institut für Theoretische Physik, Westfälische Wilhelms-Universität Münster,\\Wilhelm-Klemm-Str. 9, 48149 Münster, Germany}

\emailAdd{thede.deboer@mpi-hd.mpg.de}
\emailAdd{michael.klasen@uni-muenster.de}

\abstract{We systematically study and classify scotogenic models with a local U(1) gauge symmetry. These models give rise to radiative neutrino masses and a stable dark matter candidate, but avoid the theoretical problems of global and discrete symmetries. We restrict the dark sector particle content to up to four scalar or fermionic SU(2) singlets, doublets or triplets and use theoretical arguments based on anomaly freedom, Lorentz and gauge symmetry to find all possible charge assignments of these particles. The U(1) symmetry can be broken by a new Higgs boson to a residual discrete symmetry, that still stabilizes the dark matter candidate. We list the particle content and charge assignments of all non-equivalent models. Specific examples in our class of models that have been studied previously in the literature are the U(1)$_D$ scotogenic and singlet-triplet scalar models breaking to $Z_2$. We also briefly discuss the new phenomenological aspects of our model arising from the presence of a new massless dark photon or massive $Z'$ boson as well as the additional Higgs boson.}

\begin{document}
\maketitle
\flushbottom

\section{Introduction}
\label{sec:1}

The discovery of neutrino oscillations indicates that the neutrino masses are non-zero, but extremely small \cite{Zyla:2020zbs,Esteban:2020cvm}. While the mass generation of most Standard Model (SM) particles can be understood within the SM Higgs mechanism following the discovery of a scalar boson of mass 125 GeV by ATLAS \cite{ATLAS:2012yve} and CMS \cite{CMS:2012qbp} at the LHC, the SM can not convincingly explain this smallness of the neutrino masses. On the other hand, cosmological observations at many different length scales provide ample evidence for the existence of cold Dark Matter (DM) in the Universe \cite{Zyla:2020zbs,Klasen:2015uma}, although its exact nature remains an open question. In radiative seesaw models, these two open ends are intimately connected. There, neutrinos interact with the Higgs boson indirectly, and neutrino masses are generated at the loop level \cite{Bonnet:2012kz,Cai:2017jrq}. The particles in the loop can then contain one or more DM candidates in the form of Weakly Interacting Massive Particles (WIMPs) \cite{Restrepo:2013aga}. The scotogenic model proposed by Ernest Ma \cite{Ma:2006km} (see also Ref.\ \cite{Tao:1996vb}), which connects a second, inert Higgs doublet with additional, sterile neutrinos, is a well-known example. It has been widely studied in the literature in its original formulation \cite{Klasen:2013jpa,Vicente:2014wga,Toma:2013zsa,deBoer:2020yyw,deBoer:2021pon} and more generalized versions \cite{Escribano:2020iqq}. Radiative seesaw models containing other multiplets have also been studied \cite{Esch:2016jyx,Esch:2018ccs,Fiaschi:2018rky,deBoer:2021xjs}.

In order to prevent a tree-level seesaw contribution and to guarantee DM stability, a discrete $Z_2$ symmetry is often imposed on the new multiplets. This particular choice is, however, not very well justified theoretically. A similar role can, e.g., be played by a $Z_3$ \cite{Ma:2007gq,Belanger:2012zr,Aoki:2014cja}, $Z_4$ \cite{BenTov:2012tg} or a higher $Z_n$ \cite{Belanger:2014bga} symmetry, which can even be linked to an $A_4$ neutrino family symmetry \cite{Ma:2008ym,BenTov:2012tg,Ma:2015roa}. In addition, there exist several theoretical arguments why discrete symmetries should have a dynamical origin. One motivation is that the spontaneous breaking of discrete symmetries to non-trivial subgroups leads to the formation of domain walls, which have severe cosmological problems \cite{Zeldovich:1974uw}. These can be avoided in models with a global U(1) symmetry, which have indeed been used to generate neutrino masses at the one- \cite{Arhrib:2015dez,Fiaschi:2019evv} and two-loop \cite{Lindner:2011it,Bonilla:2016diq} level. When the global U(1) is spontaneously broken, a discrete $Z_n$ remains \cite{AristizabalSierra:2014irc}. However, a second reason why global symmetries should be of dynamical origin is that they are violated by quantum gravity effects \cite{Ali:2020znq}. This argument applies even to continuous global symmetries \cite{Bekenstein:1972ky,Abbott:1989jw,Kallosh:1995hi}.

It is therefore the aim of this paper to systematically study and classify scotogenic models, where the tree-level seesaw mechanism is forbidden by and the stability of DM is obtained from a {\em local} U(1) symmetry. Following the principle of Occham's razor, we focus on models where the neutrino masses are generated at the one-loop level and which are anomaly-free without the postulation of additional particles. In Secs.\ \ref{sec:2} and \ref{sec:3} we give detailed arguments on how our models are constructed. Our main result, i.e. the list of models is given in table format in Sec.\ \ref{sec:3}. The corresponding classification of models with a discrete $Z_2$ symmetry and up to four dark multiplets has been performed in Ref.\ \cite{Restrepo:2013aga}. Specific examples in our class of models that have been studied previously in the literature are the U(1)$_D$ scotogenic \cite{Ma:2013yga} and singlet-triplet scalar \cite{Brdar:2013iea} models breaking to $Z_2$. Models that rely on assumptions different from ours are, of course, also possible, e.g.\ models with a gauged U(1)$_{B-L}$ symmetry breaking to $Z_n$ \cite{Li:2010rb,Ho:2016aye,Bonilla:2018ynb,Calle:2018ovc,Jana:2019mgj,Dasgupta:2019rmf,Kanemura:2011vm,Seto:2016pks} or with a gauged U(1)$_L$ symmetry breaking to $Z_3$ \cite{Ma:2019coj,Ma:2021fre}, but they are either not anomaly-free or require a larger number of new multiplets. A study similar to ours for Dirac neutrinos has been performed in Ref.\ \cite{Bernal:2021ezl}.

The addition of a local U(1) symmetry gives rise to a new vector boson, which is a massless dark photon \cite{Ackerman:2008kmp} in case of an unbroken symmetry or a massive $Z'$ boson \cite{Langacker:2008yv,Fuks:2007gk,Jezo:2014kla,Jezo:2014wra,Bonciani:2015hgv,Altakach:2020ugg,Buras:2021btx,Farzan:2017xzy,Klasen:2016qux,Okada:2016gsh,Camargo:2019mml} if the U$(1)$ is spontaneously broken. Even with the SM fields uncharged under the new U$(1)$, the new gauge boson can be searched for via the kinetic mixing portal. Giving mass to the $Z'$ boson by the Higgs mechanism also gives rise to a new scalar boson and modified Higgs couplings. In Sec.\ \ref{sec:4} we briefly discuss the main aspects of the phenomenology. Finally we draw our conclusions in Sec.\ \ref{sec:5}.

\section{Theoretical considerations}
\label{sec:2}

\begin{figure}
\centering
\includegraphics[width=0.8\textwidth]{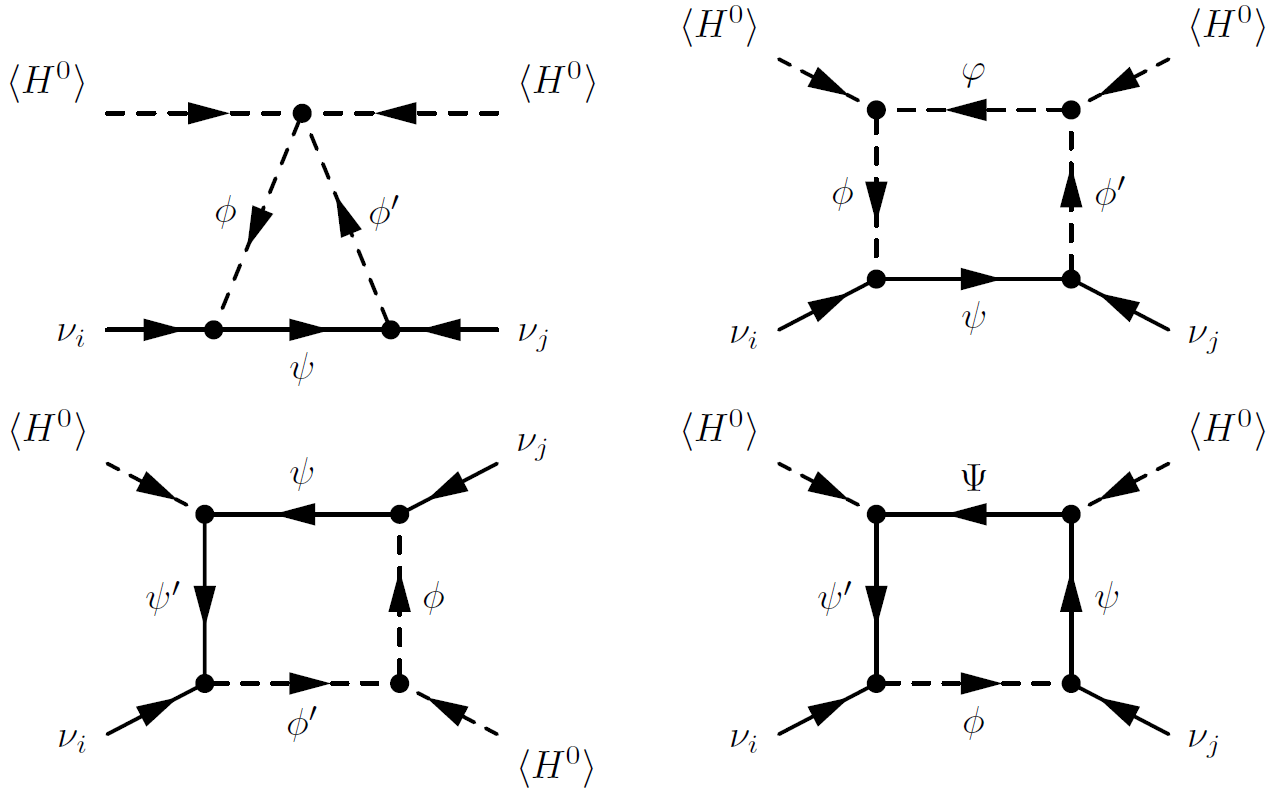}
\caption{Topologies T3, T1-1, T1-2, T1-3 of scotogenic neutrino mass generation at one loop \cite{Restrepo:2013aga}.}
\label{fig:1}
\end{figure}

Majorana neutrino masses can be generated through the effective $d=5$ Weinberg operator \cite{Weinberg:1979sa}
\begin{align}\label{eq:weinberg_op}
 \mathcal{L}\supset -\frac{c_{\alpha\beta}}{\Lambda}\left(L^\alpha H\right)\left(L^\beta H\right) +{\rm h.c.},
\end{align}
where $L^{\alpha,\beta}$ is the left-handed Weyl fermion denoting the SM lepton doublet of flavor $\alpha,\beta$, $H$ is the SM Higgs doublet, $\Lambda$ is the mass scale of the new particles and $c_{\alpha\beta}$ is obtained by integrating out the new fields. After electroweak symmetry breaking (EWSB), this operator turns into a Majorana mass term for the SM neutrinos
\begin{align}
 \mathcal{L}\supset -\frac{c_{\alpha\beta} v_H^2}{2\Lambda}\nu_L^\alpha\nu_L^\beta+{\rm h.c.},
\end{align}
which is suppressed by the scale of new physics. Tree level realizations of the Weinberg operator are the well known type I-III seesaw mechanisms. Radiative seesaw models are the one (or more) loop realizations \cite{Bonnet:2012kz,Cai:2017jrq}, which naturally suppress the neutrino mass and decrease the scale of new physics from the GUT scale to the TeV scale.

The possible realizations of this operator at one loop with a viable DM candidate have been systematically classified in Ref.\ \cite{Restrepo:2013aga} with the assumptions that the number of new fields is $\leq 4$ and that they are singlets under SU(3)$_C$ and singlets, doublets or triplets under SU(2)$_L$. A discrete $Z_2$ symmetry was imposed to prevent tree-level neutrino masses and to stabilize the DM. The four different possible topologies are shown in Fig.\ \ref{fig:1}.

The objective of this work is to replace the $Z_2$ symmetry by a gauged U$(1)_X$ symmetry, which may or may not be broken. In order to keep the models minimal, we increase the particle content of Refs.\ \cite{Bonnet:2012kz,Restrepo:2013aga} only by the neutral extra gauge boson $Z'$ and, for a spontaneously broken U(1)$_X$, by an additional scalar field $\zeta$ of charge $X_\zeta$. This scalar must be a singlet under the SM gauge group, as it should only break the new gauge symmetry. 

\subsection{Gauge invariance of the Weinberg operator}

Several theoretical constraints must be taken into account when using a U(1)$_X$ gauge symmetry to stabilize DM in radiative seesaw models. The first condition is that the Weinberg operator should be allowed and gauge invariant under the full gauge group SU(3)$_C\times$SU(2)$_L\times$U(1)$_Y\times$U(1)$_X$. This holds even if U(1)$_X$ is spontaneously broken (see below) and implies in particular for the charges of the left-handed SM leptons $X_L$ and of the SM Higgs boson $X_H$ under the new gauge group U(1)$_X$ that
\begin{align}
	X_L+X_H&=0.
\label{eq:2.3}
\end{align}
Here we assume that the new symmetry does not distinguish between different generations.

\subsection{Gauge invariance of the SM Yukawa interactions}

Second, the SM Yukawa interactions
\begin{align}
	\mathcal{L}\supset -\frac{y_d}{\sqrt{2}}H^\dagger Q d_R^c-\frac{y_u}{\sqrt{2}}H Q u_R^c-\frac{y_e}{\sqrt{2}}H^\dagger L e_R^c + {\rm h.c.}
\end{align}
should remain gauge invariant after adding the U(1)$_X$ symmetry. This implies for the U(1)$_X$ charges of the left-handed quark and lepton doublets $Q,L$ and the right-handed quark and lepton singlets $u_R,d_R,e_R$ that
\begin{align}
	X_Q-X_H+X_{d_R^c}&=0,\\
	X_Q+X_H+X_{u_R^c}&=0,\\
	X_L-X_H+X_{e_R^c}&=0.
\end{align}
Together with Eq.\ (\ref{eq:2.3}), this simplifies to
\begin{align}
	X_{d_R^c}&=-X_Q-X_L,\label{eq:yukawa_cond1}\\
	X_{u_R^c}&=-X_Q+X_L,\label{eq:yukawa_cond2}\\
	X_{e_R^c}&=-2X_L.\label{eq:yukawa_cond3}
\end{align}

\subsection{Anomaly freedom}

Third, we require our models to be anomaly-free with the given particle content. The conditions for the gauge (and gravity) anomalies to cancel are listed in Tab.\ \ref{tab:AnomalyConditions}.
\begin{table}
\caption{Conditions for gauge anomaly cancellation. $Y_\psi$ denotes the SM U(1)$_Y$ hypercharge, $X_\psi$ the new U(1)$_X$ charge of the left-handed Weyl fermion $\psi$. For SU(3)$_C$ we consider singlets $1_3$ and triplets $3_3$, while for SU(2)$_L$ we consider singlets $1_2$, doublets $2_2$ and triplets $3_2$. The sums run over all components.}
\label{tab:AnomalyConditions}
\centering
\begin{tabular}{|c|c|}
		\hline
		Anomaly & Constraint   \\
		\hline
		SU$(3)_C^2\times$U$(1)_Y$ & $\sum_{\psi\in3_3} Y_\psi=0$\\
		SU$(3)_C^2\times$U$(1)_X$ & $\sum_{\psi\in3_3} X_\psi=0$\\
		SU$(2)_L^2\times$U$(1)_Y$ & $\sum_{\psi\in2_2} Y_\psi+4\sum_{\psi\in3_2} Y_\psi=0$\\
		SU$(2)_L^2\times$U$(1)_X$ & $\sum_{\psi\in2_2} X_\psi+4\sum_{\psi\in3_2} X_\psi=0$\\
		U$(1)_Y^3$ & $\sum_\psi Y_\psi^3=0$ \\
		U$(1)_X^3$ & $\sum_\psi X_\psi^3=0$ \\
		U$(1)_Y^2\times$U$(1)_X$ & $\sum_\psi Y_\psi^2X_\psi=0$ \\
		U$(1)_X^2\times$U$(1)_Y$ & $\sum_\psi X_\psi^2Y_\psi=0$ \\
		grav$^2\times$U$(1)_Y$ & $\sum_{\psi} Y_\psi=0$ \\
		grav$^2\times$U$(1)_X$ & $\sum_{\psi} X_\psi=0$ \\
		\hline
\end{tabular}
\end{table}
Using Eqs.\ (\ref{eq:yukawa_cond1})-(\ref{eq:yukawa_cond3}), the SM contributions to the new gauge anomalies can be expressed as:
\begin{align}
	\text{SU}(3)_C^2\times\text{U}(1)_X&:\quad 0,\label{eq:2.11}\\ 
	\text{SU}(2)_L^2\times\text{U}(1)_X&:\quad 6\left[3X_Q+X_L\right],\label{eq:2.12}\\ 
	\text{U}(1)_X^3&:\quad -6X_L^2\left[3X_Q+X_L\right],\label{eq:2.13}\\ 
	\text{U}(1)_Y^2\times\text{U}(1)_X&:\quad -\frac{3}{2}\left[3X_Q+X_L\right],\label{eq:2.14}\\ 
	\text{U}(1)_X^2\times\text{U}(1)_Y&:\quad 6X_L\left[3X_Q+X_L\right],\label{eq:2.15}\\ 
	\text{grav}^2\times\text{U}(1)_X&:\quad 0. \label{eq:2.16} 
\end{align}
They must therefore either vanish or be canceled by contributions of the new fields. In addition, the Witten anomaly must cancel, which is the case for an even number of fermion doublets.

\subsection{New fermions must be vectorlike}

Since contributions to the anomalies from vector-like fermions cancel among their left- and right-handed components, only fermions that are not vector-like must be considered in more detail. As we will see, anomaly cancellation requires all our new fermions to be vector-like.\\

\noindent \underline{Models with one new fermion (T3, T1-1)}\\

If there is a single new fermion $\psi$, that is not a priori part of a vector-like fermion, $\psi$ must be made vector-like. From the Witten anomaly it is clear that if $\psi$ is a doublet, it must be made vector-like as we must add a second doublet. The two Weyl fermions then combine to a single Dirac fermion. If $\psi$ is a singlet or a triplet, the anomalies associated with the SM hypercharge must cancel. As the SM and other new vector-like fermions do not contribute to the hypercharge anomalies, $\psi$ must either be vector-like or have zero hypercharge. However, if a singlet or triplet has zero hypercharge, it must have a non-zero U$(1)_X$ charge, as otherwise seesaw types I or III are possible. This holds also if the SM neutrino is charged under U$(1)_X$, as $X_L=-X_H$. In this case the $\text{grav}^2\times\text{U}(1)_X$ anomaly, which has no contributions from the SM (cf.\ Eq.\ (\ref{eq:2.16})), must cancel. This is only possible if $\psi$ is vector-like.\\

\noindent \underline{Models with two new fermions (T1-2)}\\

If there are two (or more) new fermions in the neutrino loop, at least one of them must be a doublet and at least one must be a singlet or triplet, since they must couple to the SM Higgs doublet in a gauge-invariant way. As the doublet must be vector-like to cancel the Witten anomaly, it follows from the arguments made above that the second fermion must also be vector-like. \\

\noindent \underline{Models with three new fermions (T1-3)}\\

Again, at least one fermion must be a (vector-like) doublet in order to couple to the SM Higgs boson. Several cases have to be distinguished. (i) In the neutrino loop, three doublets can not couple to the two SM Higgs bosons in a gauge-invariant way. (ii) If there are two vector-like doublets, the third singlet/triplet must also be vector-like (see above). (iii) If the two doublets are a priori not vector-like, they must have opposite hypercharge to cancel the grav$^2\times$U$(1)_Y$ and SU(2)$_L^2\times$U$(1)_Y$ anomalies. These conditions also impose that the third singlet/triplet must be vector-like or have zero hypercharge. With this result, the U(1)$_X$ anomaly conditions then require the two doublets to be vector-like or have opposite U(1)$_X$ charge, so that they can be identified with each other. It also follows from above that the third singlet/triplet must also be vector-like. (iv) If one fermion is a (vector-like) doublet and the other two are singlets or triplets, the latter must be vector-like, identified with each other or have zero hypercharge to cancel the SM hypercharge anomalies. Then there is no BSM contribution to the $\text{U}(1)_X^2\times\text{U}(1)_Y$ anomaly. Since Eq.\ (\ref{eq:2.15}) implies that the SM contributions must cancel, also the BSM contributions must cancel among themselves. As singlets and triplets with zero hypercharge must have a U$(1)_X$ charge in order to avoid seesaw types I and III (see above), one finds that they must be vector-like or identified with each other.\\

To sum up, we find that all new fermions must be vector-like or be combined with another fermion to a vector-like fermion, whose components can form a Dirac mass term.

\subsection{SM fermions must be neutral}

Since all new fermions must be vector-like, the expressions in Eqs.\ (\ref{eq:2.12})-(\ref{eq:2.15}) must all vanish. This is only possible with the two orthogonal solutions that the new charges of the SM particles are either proportional to their SM hypercharges or the $B-L$ charges \cite{Schwartz:2013pla}
\begin{align}
	X_L=-X_{e_R^c}=-X_{\nu_R^c}=-1,\qquad X_Q=-X_{u_R^c}=-X_{d_R^c}=\frac{1}{3},
\end{align}
where $X_{\nu_R^c}$ is the new charge of a right-handed neutrino. Since we do not allow for a right-handed neutrino to avoid Dirac neutrino masses and the tree-level seesaw mechanism, this possibility is ruled out. Note also that the Weinberg operator violates the $B-L$ symmetry.  

The only possibility is therefore to assign the SM particles a U(1)$_X$ charge proportional to their hypercharge, which has a similar effect as gauge kinetic mixing, since the SM particles then couple also to the new gauge boson. Since for all topologies the SM hypercharge is conserved and $\zeta$ has zero hypercharge, the charges of the new particles running in the neutrino loop are shifted at each vertex by
\begin{align}
    X_{\rm out}= X_{\rm in} + \lambda Y_{\rm SM}.
\end{align}
One can then change the basis of the two U(1) groups and shift the U$(1)_X$ charges of the SM particles by $-\lambda Y_{\rm SM}$, which leaves us with the case that the SM is uncharged under U$(1)_X$, whereas all new fields keep their original charges.

\subsection{New scalars and fermions must be charged}

The stability of DM requires that neither the DM nor any other new particle in the neutrino loop is uncharged under the new U(1)$_X$ symmetry. Otherwise the DM particle could decay into SM particles either directly or through a diagram resulting from cutting the loop at the propagators of the DM candidate and the uncharged particle.

The $Z'$ gauge boson of the abelian group U(1)$_X$ remains of course uncharged, whereas the new scalar $\zeta$ must be charged in order to break the new symmetry spontaneously.

\subsection{Unbroken U(1)$_X$ symmetry}

Since the SM is uncharged, all new particles in the neutrino loop can have the same U$(1)_X$ charge. This is the case for an unbroken U(1)$_X$ symmetry, but also if U(1)$_X$ is spontaneously broken through a vacuum expectation value (VEV) $v_\zeta$ much smaller than the scale of new physics $\Lambda$. An even number of new scalars $\zeta$ can then in principle couple to the neutrino loop and lead to the higher-dimensional effective operators
\begin{align}
	\mathcal{L}\supset -\frac{c_{\alpha\beta}}{\Lambda^{1+2n}}\left(L^\alpha H\right)\left(L^\beta H\right)|\zeta|^{2n}+{\rm h.c.}\label{eq:NeutrinoMassOperator}
\end{align}
However, when $\zeta$ obtains a VEV $v_\zeta\ll\Lambda$, these contributions are suppressed by at least one power of $v_\zeta^2/\Lambda^2$, and in the limit $v_\zeta\to0$ only the Weinberg operator is relevant for neutrino mass generation.

\subsection{Broken U(1)$_X$ symmetry}

In contrast, when $v_\zeta\approx\Lambda$, the higher-dimensional effective operators are of equal importance as the Weinberg operator and must also be taken into account. In this case we can still use the results of Ref.\ \cite{Restrepo:2013aga} as a complete classification.

However, mass dimension three vertices appearing in the topologies T1-1 and T1-2 after the breaking of the U(1)$_X$ symmetry can violate the U$(1)_X$ charge by one unit of $X_\zeta$ through the vertex $H \phi \phi' \zeta$ or similar terms with conjugate fields allowed by gauge (and Lorentz) invariance. This may lead to additional charge assignments.

Propagators (appearing in all topologies) can also violate the U$(1)_X$ charge by one unit of $X_\zeta$ for fermions and by one or two units for scalars through the vertices $\psi\psi'\zeta$, $\phi\phi'\zeta$ and $\phi\phi'\zeta\zeta$. The fields $\psi(\phi)$ and $\psi'(\phi')$ must then be in conjugate SU(3)$_C\times$SU(2)$_L\times$U(1)$_Y$ representations. For zero hypercharge, one may also have $\psi=\psi'$ and $\phi=\phi'$, which implies $X_{\psi/\phi}=X_\zeta/2$. In the following, we set $X_\zeta = 2$ without loss of generality, since one can always rescale the U$(1)_X$ gauge coupling.

Note that in general the new scalars $\phi$ and fermions $\psi$ can not have the same charge as $\zeta$. For scalars $\phi$, vertices like $H\phi\zeta$ for doublets or $H\phi\zeta H$ for singlets/triplets would otherwise induce DM mixing with the SM Higgs or DM decay after U(1)$_X$ symmetry breaking. This also implies that the term $\phi\phi\zeta\zeta$ is not allowed. For fermions $\psi$, the fermionic vertices in the neutrino loops always imply a scalar with the same charge, which brings us back to the argument for scalars.

\subsection{Residual global symmetry}
\label{sec:2.9}
Before U$(1)_X$ breaking, the Lagrangian is invariant under the local gauge transformation $e^{i\alpha(x) X_{\phi,\psi,\zeta}}$. For fixed $\alpha={2\pi/X_\zeta}$, $\zeta$ transforms trivially, so that after U$(1)_X$ breaking the Lagrangian is still invariant under the global transformation $e^{i2\pi {X_{\phi,\psi}/X_\zeta}}$. For a fixed ratio $X_{\phi}/X_\zeta=r$, the charges of the other fields vary only by $n\in Z$ units of $X_\zeta$. It is obvious that this variation does not affect the global invariance of the Lagrangian, since $e^{i2\pi (r\pm n)}=e^{i2\pi r}$. The ratio $r$, however, does, so that for $r={1\over2},{1\over3},{1\over4}$ only a global $Z_{2,3,4}$ symmetry remains. Depending on the model, the residual symmetry can also be larger, in particular a global $U(1)_X$. The models with a residual $Z_2$ symmetry are similar to those in Ref.\ \cite{Restrepo:2013aga} apart from the new gauge and Higgs bosons and the fact that all scalars are complex and all fermions vector-like. The residual symmetry still stabilizes DM and prevents a tree-level seesaw mechanism.

\section{One-loop scotogenic models with a local U(1) symmetry}
\label{sec:3}

We will now discuss the possible charge assignments in our models. Recall that for all topologies and all choices of SU(2)$_L$ multiplets and SM hypercharge parameter $\alpha$, we can assign the same U$(1)_X$ charge $\beta$ to the new scalar and fermion fields. For a broken U(1)$_X$ with $v_\zeta\approx\Lambda$, additional possibilities are found by allowing U$(1)_X$ violation in vertices and propagators, leading to particle mixing. These cases are discussed individually. At the end of each subsection, we give a list of non-equivalent models of the respective topology. Models that are inconsistent with direct detection bounds have been omitted.

\subsection{T1-1}

First, we can always have the assignment
\begin{align}
	X_\varphi=X_{\phi'}=X_\psi=X_\phi=\beta.
\end{align}
Second, in this topology both vertices that couple to the SM Higgs boson can violate U$(1)_X$ through an additional coupling of $\zeta$. We can therefore also have
\begin{align}
	X_\varphi\pm X_\zeta=X_{\phi'}=X_\psi=X_\phi=\beta.
\end{align}
Third, if $\zeta$ couples to the propagators, one may find additional assignments for a subset of SM hypercharge parameters $\alpha$, which we discuss individually.

\paragraph{\boldmath $\alpha=1$}
$\phi'$ has zero hypercharge. All new charge assignments are equivalent to those already found once one redefines $\phi'\to\left(\phi'\right)^\dagger$.

\paragraph{\boldmath $\alpha=-1$}
$\phi$ has zero hypercharge. All new charge assignments are equivalent to those already found once one redefines $\phi\to\left(\phi\right)^\dagger$.

\paragraph{\boldmath $\alpha=0$}
Both $\psi$ and $\varphi$ have zero hypercharge. One finds one new non-equivalent charge assignment given by
\begin{align}
	X_\varphi=-X_{\phi'}=X_\psi=X_\phi=\pm\frac{X_\zeta}{2}.
\end{align}
$\phi$ and $\phi'$ also have opposite hypercharge. If they are in addition in the same representation of SU$(2)_L$, mixing between $\phi$ and $\left(\phi'\right)^\dagger$ can be induced by the breaking of U$(1)_X$. This allows for the non-equivalent charge assignments
\begin{align}
    X_\varphi=X_{\phi'}\pm2X_\zeta=X_\psi=X_\phi=\pm\frac{X_\zeta}{2},\\
    X_\varphi=X_{\phi'}\mp X_\zeta=X_\psi=X_\phi=\pm\frac{X_\zeta}{2},\\
    X_\varphi\mp X_\zeta=X_{\phi'}\mp X_\zeta=X_\psi=X_\phi=\pm\frac{X_\zeta}{2}.
\end{align}
If $\phi$ and $\phi'$ are in the same representation of SU$(2)_L$ and have opposite U$(1)_X$ charge, we can identify them with each other by defining $\phi'=\phi^\dagger$. However, in order to have at least two massive neutrinos (see below), there must be then two generations of either $\psi$ or $\phi$. The latter case is equivalent to the case where the fields are not identified with each other.\\

\begin{table}[p]
\caption{Non-equivalent models of topology T1-1. $X_\zeta$ is normalized to 2.}
\label{tab:T1-1}
\centering
\begin{tabular}{|c|c||c|c|c|c||c|c|c|c|}
\hline
	Model & $\alpha$&  $\varphi$ & $\phi'$ & $\psi$ & $\phi$ & $X_\varphi$ & $X_{\phi'}$ & $X_\psi$ & $X_\phi$ \\\hline\hline
	T1-1-A & $   0$ & $1^{S}_{ 0}$ &  $2^{S}_{-1}$ & $1^{F}_{ 0}$ & $2^{S}_{ 1}$ & $  \beta  $ & $  \beta$ & $  \beta$ & $  \beta$ \\\hline
	T1-1-A & $   0$ & $1^{S}_{ 0}$ &  $2^{S}_{-1}$ & $1^{F}_{ 0}$ & $2^{S}_{ 1}$ & $\beta\pm2$ & $  \beta$ & $  \beta$ & $  \beta$ \\\hline
	T1-1-A & $   0$ & $1^{S}_{ 0}$ &  $2^{S}_{-1}$ & $1^{F}_{ 0}$ & $2^{S}_{ 1}$ & $    1$ & $ -1$ & $  1$ & $  1$ \\\hline
	T1-1-A & $   0$ & $1^{S}_{ 0}$ &               & $1^{F}_{ 0}$ & $2^{S}_{ 1}$ & $    1$ &       & $  1$ & $  1$ \\\hline
	T1-1-A & $   0$ & $1^{S}_{ 0}$ &  $2^{S}_{-1}$ & $1^{F}_{ 0}$ & $2^{S}_{ 1}$ & $    1$ & $ \pm3$ & $  1$ & $  1$ \\\hline
	T1-1-A & $   0$ & $1^{S}_{ 0}$ &  $2^{S}_{-1}$ & $1^{F}_{ 0}$ & $2^{S}_{ 1}$ & $    3$ & $  3$ & $  1$ & $  1$ \\\hline\hline
	T1-1-A & $\pm2$ & $1^{S}_{\pm 2}$ &  $2^{S}_{ 1,-3}$ & $1^{F}_{\pm 2}$ & $2^{S}_{ 3,-1}$ & $  \pm1 $ & $  1 $ & $  1$ & $ 1$ \\\hline
	T1-1-A & $\pm2$ & $1^{S}_{\pm 2}$ &  $2^{S}_{ 1,-3}$ & $1^{F}_{\pm 2}$ & $2^{S}_{ 3,-1}$ & $  3 $ & $  1 $ & $  1$ & $ 1$ \\\hline\hline
	T1-1-B & $   0$ & $1^{S}_{ 0}$ &  $2^{S}_{-1}$ & $3^{F}_{ 0}$ & $2^{S}_{ 1}$ & $  \beta  $ & $  \beta$ & $  \beta$ & $  \beta$ \\\hline
	T1-1-B & $   0$ & $1^{S}_{ 0}$ &  $2^{S}_{-1}$ & $3^{F}_{ 0}$ & $2^{S}_{ 1}$ & $\beta\pm2$ & $  \beta$ & $  \beta$ & $  \beta$ \\\hline
	T1-1-B & $   0$ & $1^{S}_{ 0}$ &  $2^{S}_{-1}$ & $3^{F}_{ 0}$ & $2^{S}_{ 1}$ & $    1$ & $ -1$ & $  1$ & $  1$ \\\hline
	T1-1-B & $   0$ & $1^{S}_{ 0}$ &               & $3^{F}_{ 0}$ & $2^{S}_{ 1}$ & $    1$ &       & $  1$ & $  1$ \\\hline
	T1-1-B & $   0$ & $1^{S}_{ 0}$ &  $2^{S}_{-1}$ & $3^{F}_{ 0}$ & $2^{S}_{ 1}$ & $    1$ & $ \pm3$ & $  1$ & $  1$ \\\hline
	T1-1-B & $   0$ & $1^{S}_{ 0}$ &  $2^{S}_{-1}$ & $3^{F}_{ 0}$ & $2^{S}_{ 1}$ & $    3$ & $  3$ & $  1$ & $  1$ \\\hline\hline
	T1-1-B & $\pm2$ & $1^{S}_{\pm 2}$ &  $2^{S}_{ 1,-3}$ & $3^{F}_{\pm 2}$ & $2^{S}_{ 3,-1}$ & $  \pm1 $ & $  1 $ & $  1$ & $ 1$ \\\hline
	T1-1-B & $\pm2$ & $1^{S}_{\pm 2}$ &  $2^{S}_{ 1,-3}$ & $3^{F}_{\pm 2}$ & $2^{S}_{ 3,-1}$ & $  3 $ & $  1 $ & $  1$ & $ 1$ \\\hline\hline
	T1-1-C & $\pm1$ & $2^{S}_{\pm 1}$ &  $1^{S}_{ 0,-2}$ & $2^{F}_{\pm 1}$ & $1^{S}_{ 2,0}$ & $  \beta  $ & $  \beta$ & $  \beta$ & $  \beta$ \\\hline
	T1-1-C & $\pm1$ & $2^{S}_{\pm 1}$ &  $1^{S}_{ 0,-2}$ & $2^{F}_{\pm 1}$ & $1^{S}_{ 2,0}$ & $\beta\pm2$ & $  \beta$ & $  \beta$ & $  \beta$ \\\hline\hline
	T1-1-D & $   1$ & $2^{S}_{ 1}$ &  $1^{S}_{ 0}$ & $2^{F}_{ 1}$ & $3^{S}_{ 2}$ & $  \beta  $ & $  \beta$ & $  \beta$ & $  \beta$ \\\hline
	T1-1-D & $   1$ & $2^{S}_{ 1}$ &  $1^{S}_{ 0}$ & $2^{F}_{ 1}$ & $3^{S}_{ 2}$ & $\beta\pm2$ & $  \beta$ & $  \beta$ & $  \beta$ \\\hline\hline
	T1-1-D & $  -1$ & $2^{S}_{-1}$ &  $1^{S}_{-2}$ & $2^{F}_{-1}$ & $3^{S}_{ 0}$ & $  \beta  $ & $  \beta$ & $  \beta$ & $  \beta$ \\\hline
	T1-1-D & $  -1$ & $2^{S}_{-1}$ &  $1^{S}_{-2}$ & $2^{F}_{-1}$ & $3^{S}_{ 0}$ & $\beta\pm2$ & $  \beta$ & $  \beta$ & $  \beta$ \\\hline\hline
	T1-1-F & $\pm1$ & $2^{S}_{\pm 1}$ &  $3^{S}_{ 0,-2}$ & $2^{F}_{\pm 1}$ & $3^{S}_{ 2,0}$ & $  \beta  $ & $  \beta$ & $  \beta$ & $  \beta$ \\\hline
	T1-1-F & $\pm1$ & $2^{S}_{\pm 1}$ &  $3^{S}_{ 0,-2}$ & $2^{F}_{\pm 1}$ & $3^{S}_{ 2,0}$ & $\beta\pm2$ & $  \beta$ & $  \beta$ & $  \beta$ \\\hline\hline
	T1-1-G & $   0$ & $3^{S}_{ 0}$ &  $2^{S}_{-1}$ & $1^{F}_{ 0}$ & $2^{S}_{ 1}$ & $  \beta  $ & $  \beta$ & $  \beta$ & $  \beta$ \\\hline
	T1-1-G & $   0$ & $3^{S}_{ 0}$ &  $2^{S}_{-1}$ & $1^{F}_{ 0}$ & $2^{S}_{ 1}$ & $\beta\pm2$ & $  \beta$ & $  \beta$ & $  \beta$ \\\hline
	T1-1-G & $   0$ & $3^{S}_{ 0}$ &  $2^{S}_{-1}$ & $1^{F}_{ 0}$ & $2^{S}_{ 1}$ & $    1$ & $ -1$ & $  1$ & $  1$ \\\hline
	T1-1-G & $   0$ & $3^{S}_{ 0}$ &               & $1^{F}_{ 0}$ & $2^{S}_{ 1}$ & $    1$ &       & $  1$ & $  1$ \\\hline
	T1-1-G & $   0$ & $3^{S}_{ 0}$ &  $2^{S}_{-1}$ & $1^{F}_{ 0}$ & $2^{S}_{ 1}$ & $    1$ & $ \pm3$ & $  1$ & $  1$ \\\hline
	T1-1-G & $   0$ & $3^{S}_{ 0}$ &  $2^{S}_{-1}$ & $1^{F}_{ 0}$ & $2^{S}_{ 1}$ & $    3$ & $  3$ & $  1$ & $  1$ \\\hline\hline
	T1-1-G & $\pm2$ & $3^{S}_{\pm 2}$ &  $2^{S}_{ 1,-3}$ & $1^{F}_{\pm 2}$ & $2^{S}_{ 3,-1}$ & $  \pm 1 $ & $  1$ & $  1$ & $  1$ \\\hline
	T1-1-G & $\pm2$ & $3^{S}_{\pm 2}$ &  $2^{S}_{ 1,-3}$ & $1^{F}_{\pm 2}$ & $2^{S}_{ 3,-1}$ & $3$ & $  1$ & $  1$ & $  1$ \\\hline\hline
	T1-1-H & $   0$ & $3^{S}_{ 0}$ &  $2^{S}_{-1}$ & $3^{F}_{ 0}$ & $2^{S}_{ 1}$ & $  \beta  $ & $  \beta$ & $  \beta$ & $  \beta$ \\\hline           
	T1-1-H & $   0$ & $3^{S}_{ 0}$ &  $2^{S}_{-1}$ & $3^{F}_{ 0}$ & $2^{S}_{ 1}$ & $\beta\pm2$ & $  \beta$ & $  \beta$ & $  \beta$ \\\hline           
	T1-1-H & $   0$ & $3^{S}_{ 0}$ &  $2^{S}_{-1}$ & $3^{F}_{ 0}$ & $2^{S}_{ 1}$ & $    1$ & $ -1$ & $  1$ & $  1$ \\\hline           
	T1-1-H & $   0$ & $3^{S}_{ 0}$ &               & $3^{F}_{ 0}$ & $2^{S}_{ 1}$ & $    1$ &       & $  1$ & $  1$ \\\hline           
	T1-1-H & $   0$ & $3^{S}_{ 0}$ &  $2^{S}_{-1}$ & $3^{F}_{ 0}$ & $2^{S}_{ 1}$ & $    1$ & $ \pm3$ & $  1$ & $  1$ \\\hline
	T1-1-H & $   0$ & $3^{S}_{ 0}$ &  $2^{S}_{-1}$ & $3^{F}_{ 0}$ & $2^{S}_{ 1}$ & $    3$ & $  3$ & $  1$ & $  1$ \\\hline\hline
	T1-1-H & $\pm2$ & $3^{S}_{\pm 2}$ &  $2^{S}_{ 1,-3}$ & $3^{F}_{\pm 2}$ & $2^{S}_{ 3,-1}$ & $  \pm 1 $ & $  1$ & $  1$ & $  1$ \\\hline
	T1-1-H & $\pm2$ & $3^{S}_{\pm 2}$ &  $2^{S}_{ 1,-3}$ & $3^{F}_{\pm 2}$ & $2^{S}_{ 3,-1}$ & $3$ & $  1$ & $  1$ & $  1$ \\\hline
\end{tabular}
\end{table}

All non-equivalent models of topology T1-1 are listed in Tab.\ \ref{tab:T1-1}. As in Ref.\ \cite{Restrepo:2013aga}, the field content is denoted as $L_Y^{\mathcal{L}}$, where $L$ is the type of SU(2)$_L$ multiplet (1 for singlet, 2 for doublet, 3 for triplet), $\mathcal{L}$ denotes scalars ($S$) or fermions ($F$), and $Y\equiv 2(Q-I_3)$ is the hypercharge. Since $X_\zeta$ is set to 2, the parameter $\beta$ can not take the values $0,\pm2$. Models with scalar doublets of charge $\pm4$ are also not allowed, since the vertex $H\phi\zeta\zeta$ would then induce mixing of the new scalar with the Higgs boson and make DM unstable. 

For dark matter consisting of scalar doublets, there needs to be a mass splitting between the CP-odd and CP-even components in order to avoid direct detection limits. Therefore for the models T1-1-A ($\alpha=\pm2$),  T1-1-B ($\alpha=\pm2$),  T1-1-G ($\alpha=\pm2$) and  T1-1-H ($\alpha=\pm2$) only the charge assignment that leads to a residual symmetry of $Z_2$ is allowed. See Sec.\ \ref{sec:4.4} for details.

\subsection{T1-2}

First, we can always have the assignment
\begin{align}
	X_\psi=X_\phi=X_{\phi'}=X_{\psi'}=\beta.
\end{align}
Second, in this topology there is only one three-scalar vertex which can violate U$(1)_X$. Thus $\zeta$ must also always couple to a propagator. This leads to additional assignments for a subset of SM hypercharge parameters $\alpha$, which we discuss individually.

\paragraph{\boldmath $\alpha=-1$}
$\psi'$ and $\phi$ have zero hypercharge. This allows the new charge assignment
\begin{align}
	X_\psi=X_\phi=-X_{\phi'}=X_{\psi'}=\pm\frac{X_\zeta}{2}.
\end{align}

\paragraph{\boldmath $\alpha=0$}
$\psi$ and $\phi'$ have zero hypercharge. We find the non-equivalent charge assignment
\begin{align}
	X_\psi=-X_\phi=X_{\phi'}=X_{\psi'}=\pm\frac{X_\zeta}{2}.
\end{align}

All non-equivalent models of topology T1-2 are listed in Tab.\ \ref{tab:T1-2}. Again, the parameter $\beta\neq0,\pm2$, and the assignment $\beta=\pm4$ may yield problems with DM stability, if the SM Higgs boson mixes with a new scalar doublet. 

For the models with only scalar doublet dark matter, explicitly T1-2-A ($\alpha=-2$),  T1-2-B ($\alpha=-2$),  T1-2-D ($\alpha=1$) and  T1-2-F ($\alpha=1$), only the charge assignment that leads to a residual symmetry of $Z_2$ is allowed. See Sec.\ \ref{sec:4.4} for details.

\begin{table}
\caption{Non-equivalent models of topology T1-2. $X_\zeta$ is normalized to 2.}
\label{tab:T1-2}
\centering
\begin{tabular}{|c|c||c|c|c|c||c|c|c|c|}
\hline
	Model & $\alpha$& $\psi$ & $\phi$ & $\phi'$ & $\psi'$ & $X_\psi$ & $X_\phi$ & $X_{\phi'}$ & $X_{\psi'}$ \\\hline\hline
	T1-2-A & $   0$ & $1^{F}_{ 0}$ & $2^{S}_{ 1}$ & $1^{S}_{ 0}$ & $2^{F}_{ 1}$ & $  \beta$ & $  \beta$ & $  \beta$ & $  \beta$ \\\hline
	T1-2-A & $   0$ & $1^{F}_{ 0}$ & $2^{S}_{ 1}$ & $1^{S}_{ 0}$ & $2^{F}_{ 1}$ & $  1$ & $ -1$ & $  1$ & $  1$ \\\hline\hline
	T1-2-A & $  -2$ & $1^{F}_{-2}$ & $2^{S}_{-1}$ & $1^{S}_{-2}$ & $2^{F}_{-1}$ & $  1$ & $  1$ & $  1$ & $  1$ \\\hline\hline
	T1-2-B & $   0$ & $1^{F}_{ 0}$ & $2^{S}_{ 1}$ & $3^{S}_{ 0}$ & $2^{F}_{ 1}$ & $  \beta$ & $  \beta$ & $  \beta$ & $  \beta$ \\\hline
	T1-2-B & $   0$ & $1^{F}_{ 0}$ & $2^{S}_{ 1}$ & $3^{S}_{ 0}$ & $2^{F}_{ 1}$ & $  1$ & $ -1$ & $  1$ & $  1$ \\\hline\hline
	T1-2-B & $  -2$ & $1^{F}_{-2}$ & $2^{S}_{-1}$ & $3^{S}_{-2}$ & $2^{F}_{-1}$ & $  1$ & $  1$ & $  1$ & $  1$ \\\hline\hline
	T1-2-D & $   1$ & $2^{F}_{ 1}$ & $1^{S}_{ 2}$ & $2^{S}_{ 1}$ & $3^{F}_{ 2}$ & $  1$ & $  1$ & $  1$ & $  1$ \\\hline\hline
	T1-2-D & $  -1$ & $2^{F}_{-1}$ & $1^{S}_{ 0}$ & $2^{S}_{-1}$ & $3^{F}_{ 0}$ & $  \beta$ & $  \beta$ & $  \beta$ & $  \beta$ \\\hline
	T1-2-D & $  -1$ & $2^{F}_{-1}$ & $1^{S}_{ 0}$ & $2^{S}_{-1}$ & $3^{F}_{ 0}$ & $  1$ & $  1$ & $ -1$ & $  1$ \\\hline\hline
	T1-2-F & $   1$ & $2^{F}_{ 1}$ & $3^{S}_{ 2}$ & $2^{S}_{ 1}$ & $3^{F}_{ 2}$ & $  1$ & $  1$ & $  1$ & $  1$ \\\hline\hline
	T1-2-F & $  -1$ & $2^{F}_{-1}$ & $3^{S}_{ 0}$ & $2^{S}_{-1}$ & $3^{F}_{ 0}$ & $  \beta$ & $  \beta$ & $  \beta$ & $  \beta$ \\\hline
	T1-2-F & $  -1$ & $2^{F}_{-1}$ & $3^{S}_{ 0}$ & $2^{S}_{-1}$ & $3^{F}_{ 0}$ & $  1$ & $  1$ & $ -1$ & $  1$ \\\hline
\end{tabular}
\end{table}

\subsection{T1-3}

First, we can always have the assignment
\begin{align}
	X_\Psi=X_{\psi'}=X_\phi=X_{\psi}=\beta.
\end{align}
Second, in this topology none of the vertices can violate U$(1)_X$. Thus $\zeta$ must always couple to two propagators. This leads to additional assignments for 

\paragraph{\boldmath $\alpha=0$}
Both $\Psi$ and $\phi$ have zero hypercharge. This allows the new charge assignment
\begin{align}
	X_\Psi=-X_{\psi'}=X_\phi=X_\psi=\pm\frac{X_\zeta}{2},
\end{align}
If $\psi$ and $\psi'$ are in the same representation of SU$(2)_L$, mixing between $\psi$ and $\left(\psi'\right)^c$ can be induced by the breaking of U$(1)_X$. We find no new non equivalent charge assignment. If $\psi$ and $\psi'$ are in the same SU$(2)_L$ representation and have opposite U$(1)_X$ charge, we can combine them into a vector-like multiplet (specifically doublet) instead of making both fields vector-like. However, in order to have at least two massive neutrinos, there must then be two generations of either $\phi$ or $\psi$. The latter case is equivalent to the case where the fields are not identified with each other.

\begin{table}
\caption{Non-equivalent models of topology T1-3. $X_\zeta$ is normalized to 2.}
\label{tab:T1-3}
\centering
\begin{tabular}{|c|c||c|c|c|c||c|c|c|c|}
\hline
	Model & $\alpha$& $\Psi$ & $\psi'$ & $\phi$ & $\psi$ & $X_\Psi$ & $X_{\psi'}$ & $X_\phi$ & $X_\psi$ \\\hline\hline
	T1-3-A & $   0$ & $1^{F}_{ 0}$ & $2^{F}_{ 1}$ & $1^{S}_{ 0}$ & $2^{F}_{-1}$ & $ \beta$ & $ \beta$ & $ \beta$ & $ \beta$ \\\hline
	T1-3-A & $   0$ & $1^{F}_{ 0}$ & $2^{F}_{ 1}$ & $1^{S}_{ 0}$ & $2^{F}_{-1}$ & $ 1$ & $-1$ & $ 1$ & $ 1$ \\\hline
	T1-3-A & $   0$ & $1^{F}_{ 0}$ & $2^{F}_{ 1}$ & $1^{S}_{ 0}$ &              & $ 1$ & $-1$ & $ 1$ &      \\\hline\hline
	T1-3-B & $   0$ & $1^{F}_{ 0}$ & $2^{F}_{ 1}$ & $3^{S}_{ 0}$ & $2^{F}_{-1}$ & $ \beta$ & $ \beta$ & $ \beta$ & $ \beta$ \\\hline
	T1-3-B & $   0$ & $1^{F}_{ 0}$ & $2^{F}_{ 1}$ & $3^{S}_{ 0}$ & $2^{F}_{-1}$ & $ 1$ & $-1$ & $ 1$ & $ 1$ \\\hline
	T1-3-B & $   0$ & $1^{F}_{ 0}$ & $2^{F}_{ 1}$ & $3^{S}_{ 0}$ &              & $ 1$ & $-1$ & $ 1$ &      \\\hline\hline
	T1-3-C & $\pm1$ & $2^{F}_{\pm 1}$ & $1^{F}_{ 2,0}$ & $2^{S}_{\pm 1}$ & $1^{F}_{ 0,-2}$ & $ \beta$ & $ \beta$ & $ \beta$ & $ \beta$ \\\hline\hline
	T1-3-D & $   1$ & $2^{F}_{ 1}$ & $1^{F}_{ 2}$ & $2^{S}_{ 1}$ & $3^{F}_{ 0}$ & $ \beta$ & $ \beta$ & $ \beta$ & $ \beta$ \\\hline\hline
	T1-3-D & $  -1$ & $2^{F}_{-1}$ & $1^{F}_{ 0}$ & $2^{S}_{-1}$ & $3^{F}_{-2}$ & $ \beta$ & $ \beta$ & $ \beta$ & $ \beta$ \\\hline\hline
	T1-3-F & $\pm1$ & $2^{F}_{\pm 1}$ & $3^{F}_{ 2,0}$ & $2^{S}_{\pm 1}$ & $3^{F}_{ 0,-2}$ & $ \beta$ & $ \beta$ & $ \beta$ & $ \beta$ \\\hline\hline
	T1-3-G & $   0$ & $3^{F}_{ 0}$ & $2^{F}_{ 1}$ & $1^{S}_{ 0}$ & $2^{F}_{-1}$ & $ \beta$ & $ \beta$ & $ \beta$ & $ \beta$ \\\hline
	T1-3-G & $   0$ & $3^{F}_{ 0}$ & $2^{F}_{ 1}$ & $1^{S}_{ 0}$ & $2^{F}_{-1}$ & $ 1$ & $-1$ & $ 1$ & $ 1$ \\\hline
	T1-3-G & $   0$ & $3^{F}_{ 0}$ & $2^{F}_{ 1}$ & $1^{S}_{ 0}$ &              & $ 1$ & $-1$ & $ 1$ &      \\\hline\hline
	T1-3-H & $   0$ & $3^{F}_{ 0}$ & $2^{F}_{ 1}$ & $3^{S}_{ 0}$ & $2^{F}_{-1}$ & $ \beta$ & $ \beta$ & $ \beta$ & $ \beta$ \\\hline
	T1-3-H & $   0$ & $3^{F}_{ 0}$ & $2^{F}_{ 1}$ & $3^{S}_{ 0}$ & $2^{F}_{-1}$ & $ 1$ & $-1$ & $ 1$ & $ 1$ \\\hline
	T1-3-H & $   0$ & $3^{F}_{ 0}$ & $2^{F}_{ 1}$ & $3^{S}_{ 0}$ &              & $ 1$ & $-1$ & $ 1$ &      \\\hline\hline
\end{tabular}
\end{table}

All non-equivalent models of topology T1-3 are listed in Tab.\ \ref{tab:T1-3}. Again, the parameter $\beta\neq0,\pm2$ and in case of a scalar doublet $\neq\pm4$.

\subsection{T3}

First, we can always have the assignment
\begin{align}
	X_{\phi'}=X_\phi=X_\psi=\beta.
\end{align}
Second, in this topology none of the vertices can violate U$(1)_X$. Thus $\zeta$ must always couple to two propagators. This could lead to additional assignments for

\paragraph{\boldmath $\alpha=-1$}
$\psi$ has zero hypercharge, $\phi$ and $\phi'$ have opposite hypercharge. However, even if $\phi$ and $\phi'$ are in addition in the same representation of SU$(2)_L$ and mixing between $\phi$ and $\left(\phi'\right)^\dagger$ is induced by the breaking of U$(1)_X$, we find no additional possible charge assignments.\\

All non-equivalent models of topology T3 are listed in Tab.\ \ref{tab:T3}. Again, the parameter $\beta\neq0,\pm2$ and in case of a scalar doublet $\neq\pm4$. Concrete examples in this class are the gauged scotogenic model T3-B ($\alpha=-1,\beta=1$) \cite{Ma:2013yga,Hagedorn:2018spx} and the gauged singlet-triplet scalar and doublet fermion model T3-A ($\alpha=-2,\beta=1$) \cite{Brdar:2013iea}.

For the models with only scalar doublet dark matter, explicitly T3-B ($\alpha=1,-3 $) and T3-C ($\alpha=1,-3 $), only the charge assignment that leads to a residual symmetry of $Z_2$ is allowed. See Sec.\ \ref{sec:4.4} for details.

\begin{table}
\caption{Non-equivalent models of topology T3. $X_\zeta$ is normalized to 2.}
\label{tab:T3}
\centering
\begin{tabular}{|c|c||c|c|c||c|c|c|}
\hline
	Model & $\alpha$& $\phi'$ & $\phi$ & $\psi$ & $X_{\phi'}$ & $X_\phi$ & $X_\psi$ \\\hline\hline
	T3-A & $   0$ & $1^{S}_{ 0}$ & $3^{S}_{ 2}$ & $2^{F}_{ 1}$ & $ \beta$ & $ \beta$ & $ \beta$ \\\hline\hline
	T3-A & $  -2$ & $1^{S}_{-2}$ & $3^{S}_{ 0}$ & $2^{F}_{-1}$ & $ \beta$ & $ \beta$ & $ \beta$ \\\hline\hline
	T3-B & $1,-3$ & $2^{S}_{ 1,-3}$ & $2^{S}_{ 3,-1}$ & $1^{F}_{ 2,-2}$ & $ 1$ & $ 1$ & $ 1$ \\\hline\hline
	T3-B & $  -1$ & $2^{S}_{-1}$ & $2^{S}_{ 1}$ & $1^{F}_{ 0}$ & $ \beta$ & $ \beta$ & $ \beta$ \\\hline\hline
	T3-C & $1,-3$ & $2^{S}_{ 1,-3}$ & $2^{S}_{ 3,-1}$ & $3^{F}_{ 2,-2}$ & $ 1$ & $ 1$ & $ 1$ \\\hline\hline
	T3-C & $  -1$ & $2^{S}_{-1}$ & $2^{S}_{ 1}$ & $3^{F}_{ 0}$ & $ \beta$ & $ \beta$ & $ \beta$ \\\hline\hline
	T3-E & $0,-2$ & $3^{S}_{ 0,-2}$ & $3^{S}_{ 2,0}$ & $2^{F}_{ 1,-1}$ & $ \beta$ & $ \beta$ & $ \beta$ \\\hline
\end{tabular}
\end{table}

\section{Phenomenological considerations}
\label{sec:4}

The models proposed above give rise to a wide variety of new phenomena. By construction and most importantly, they generate (at least two) non-vanishing neutrino masses \cite{Bonnet:2012kz,Cai:2017jrq} and include at least one viable dark matter (DM) candidate \cite{Restrepo:2013aga}.
Similarly to the models with a $Z_2$ symmetry, the values of the neutrino masses, the DM particle type, relic density, direct, indirect and collider detection prospects, as well as the predicted lepton flavor violation (LFV) rates are in general quite model-dependent \cite{Klasen:2013jpa,Vicente:2014wga,Toma:2013zsa,Esch:2016jyx,Esch:2018ccs,Fiaschi:2018rky,Escribano:2020iqq,deBoer:2020yyw,deBoer:2021xjs,deBoer:2021pon}.

Replacing the $Z_2$ by a gauged U(1)$_X$ symmetry introduces a new gauge boson, which may mix with the SM U(1)$_Y$ boson through a renormalizable kinetic mixing operator \cite{Okun:1982xi,Galison:1983pa,Holdom:1985ag,Foot:1991kb,Zhang:2018fbm}. It may also obtain a mass by either the St\"uckelberg \cite{Stueckelberg:1938hvi,Kors:2004dx,Kribs:2022gri,Feldman:2007wj} or the Higgs mechanism \cite{Englert:1964et,Guralnik:1964eu,Higgs:1964pj}. In the latter case, the massive $Z'$-boson will be accompanied by a new physical scalar boson after spontaneous symmetry breaking.

The phenomenology of the dark gauge and Higgs sectors can to some extent be discussed independently of the matter sector. A massless dark photon necessarily requires another, massive DM particle. A massive dark gauge boson can in principle itself constitute DM, if it is sufficiently long-lived and its mass is sufficiently small, although we do not consider this case here \cite{Fabbrichesi:2020wbt}. Similarly, the dark Higgs boson might constitute DM, if it is lighter than the dark photon, but again we do not consider this case here \cite{Mondino:2020lsc}.

\subsection{Massless dark photons}

Replacing the $Z_2$ with a gauged U(1)$_X$ symmetry introduces a new, a priori massless gauge boson. Since only the new scalars and fermions are charged under this U(1)$_X$ group, while all SM particles are uncharged, this corresponds to the addition of a ``dark'' photon. When the fundamental Lagrangian of a model contains several abelian gauge groups, gauge kinetic mixing occurs, since the field strength tensors $F^{\mu\nu}$ are individually gauge invariant and products of different field strength tensors are allowed by gauge invariance \cite{Okun:1982xi,Galison:1983pa,Holdom:1985ag}. The Lagrangian
\begin{align}
	\mathcal{L}\supset& -\frac{1}{4} F_Y^{\mu\nu}F_{Y\mu\nu}-\frac{\epsilon}{2} F_Y^{\mu\nu}F_{X\mu\nu}-\frac{1}{4} F_X^{\mu\nu}F_{X\mu\nu}+\left(\begin{matrix}A_{X\mu} & A_{Y\mu}\end{matrix}\right)\left(\begin{matrix}g_X & 0 \\ 0 & g_Y\end{matrix}\right)\left(\begin{matrix}j_X^\mu \\ j_Y^\mu\end{matrix}\right)
\label{eq:4.1}
\end{align}
then depends on the kinetic mixing parameter $\epsilon$ and the SM and DM interaction currents $j_{X,Y}^\mu$ coupling to the gauge fields $A_{X,Y\mu}$ with strengths $g_{X,Y}$. It can be written with diagonal kinetic terms by choosing (a) either the dark photon $A_{X\mu}$ to not couple to the hypercharge current $j_Y^\mu$ or (b) the hypercharge field $A_{Y\mu}$ to not couple to the dark current $j_X^\mu$, but not both. When all contributions are included, physical processes do not depend on the choice of basis, and the kinetic mixing effects do not show up in electromagnetic and weak interactions if only SM particles are involved in the calculations \cite{Foot:1991kb,Zhang:2018fbm}.

Massless dark photons necessarily induce long-range interactions between the DM particles, which in our models must all be charged under U(1)$_X$, independent of their spin and SM quantum numbers, and come with equal numbers of positive and negative charges. Annihilations can be suppressed, if the dark matter mass is sufficiently high and the dark fine-structure constant $\alpha_X=g_X^2/(4\pi)$ is sufficiently small. Furthermore, the correct relic abundance can be obtained if the DM also couples to SM weak interactions, as long as the kinetic mixing with ordinary photons is sufficiently small. The primary limit on $\alpha_X$ then comes from the demand that the DM be effectively collisionless in galactic dynamics, which implies $\alpha_X < 10^{-3}$ for TeV-scale DM. These values are also compatible with constraints from structure formation and primordial nucleosynthesis \cite{Ackerman:2008kmp}. They have recently been updated and combined with other constraints from stars, supernovae, precision atomic physics, LFV as well as collider physics \cite{Fabbrichesi:2020wbt}. When interpreted in terms of effective field theory, LFV, primordial nucleosynthesis, star cooling and other phenomena set limits on the scale of the dimension-six operators describing DM interactions in the 1-15 TeV range \cite{Dobrescu:2004wz}. The rotation of the mixing term in Eq.\ (\ref{eq:4.1}) leaves the photon coupled to the dark sector particles with strength $\epsilon e/\sqrt{1-\epsilon^2}$. The dark sector particles are then interpreted as millicharged particles. Searches are accordingly parameterized in terms of their mass and electromagnetic coupling modulated by the mixing angle, which is then constrained to be below ${\cal O}(0.1)$ in the GeV--TeV region probed by LEP and the LHC, but orders of magnitude smaller in the regions below and above from cosmological constraints \cite{Fabbrichesi:2020wbt}.

The existence of dark radiation and a dark matter plasma may have additional effects that could significantly affect bremsstrahlung, early universe structure formation, and the Weibel instability in galactic halos. The first two result in much weaker bounds than those discussed above. The Weibel instability is an exponential magnetic-field amplification that arises, if the plasma particles have an anisotropic velocity distribution. Such anisotropies could arise, for example, during hierarchical structure formation, as subhalos merge to form more massive halos \cite{Ackerman:2008kmp}.

\subsection{Symmetry breaking and massive $Z'$ boson}
\label{sec:H_sector}
The U$(1)$ gauge symmetry can be broken through the addition of a new scalar. For this purpose we take a complex scalar $\zeta$, which we allow to develop a VEV $v_\zeta$. If $v_\zeta$ lies at the same scale as the masses of the new particles, all charge assignments that were listed in the overview in Tabs.\ \ref{tab:T1-1}-\ref{tab:T3} are possible. If $v_\zeta \ll \Lambda$, models with $n\neq0$ in the Weinberg operator Eq.\ \eqref{eq:NeutrinoMassOperator} are suppressed which might yield too small neutrino masses. Especially in the case of $v_\zeta=0$, only charge assignments where all new fields in the neutrino loop have the same U$(1)_X$ charge are possible. With the Standard Model Higgs doublet $H$ and the new scalar singlet $\zeta$ having charges of 0 and 2 under the new U$(1)_X$ group respectively, the scalar potential is given by
\begin{align}\label{eq:Higgspot}
    V = - m_H^2 (H^\dagger H) - m_\zeta^2 (\zeta^\dagger \zeta) + \frac{\lambda_H}{2}(H^\dagger H)^2 + \frac{\lambda_\zeta}{2}(\zeta^\dagger \zeta)^2 + \lambda_{H\zeta}(H^\dagger H)(\zeta^\dagger \zeta).
\end{align}
The first step is to find the minimum of the potential, which should be the case when $H$ and $\zeta$ obtain their VEVs. We denote the VEVs of these fields by
\begin{align}
    \langle H \rangle = \left(\begin{matrix} 0 \\
    \frac{v_H+h}{\sqrt{2}}
    \end{matrix}\right),
    \qquad
    \langle \zeta \rangle = \frac{v_\zeta + \sigma}{\sqrt{2}}.
\end{align}
Minimizing the potential yields
\begin{align}
	v_H^2=& \frac{2 \lambda_{H\zeta} m_\zeta^2-2\lambda_\zeta m_H^2}{\lambda_{H\zeta}^2-\lambda_\zeta \lambda_H},\\
	v_\zeta^2=& \frac{2 \lambda_H m_\zeta^2-2\lambda_{H\zeta}m_H^2}{\lambda_\zeta \lambda_H-\lambda_{H\zeta}^2}.
\end{align}
The mass matrix for the scalar fields expressed in terms of the VEVs is given by
\begin{align}
    M = \left(\begin{matrix}
    \lambda_H v_H^2 & \lambda_{H\zeta} v_H v_\zeta \\
    \lambda_{H\zeta} v_H v_\zeta & \lambda_\zeta v_\zeta^2,
    \end{matrix}\right).
\end{align}
The eigenstate dominated by the SM Higgs is the one measured at LHC \cite{ATLAS:2012yve,CMS:2012qbp}. Searches for modified Higgs couplings and a second Higgs boson constrain the scalar sector \cite{Schabinger:2005ei,Dawson:2017jja,Bechtle:2020pkv,Choi:2013zqz,Cheung:2015dta,Cheung:2015cug,ATLAS:2015ciy}. 

Now we turn to the gauge sector. In case of two abelian gauge groups, one has to introduce gauge kinetic mixing \cite{Holdom:1985ag}. Through some (non-orthogonal) basis transformations, the effect of kinetic mixing can be shifted to an off diagonal gauge coupling. 
After spontaneous symmetry breaking, the gauge bosons obtain their masses. The covariant derivative for SU$(2)_L\times$U$(1)_Y\times$U$(1)_X$ can be written as
\begin{align}
    D_\mu=\partial_\mu-i\left(g_Y A_\mu+\frac{-\epsilon g_Y}{\sqrt{1-\epsilon^2}}A'_\mu\right)Y -i \frac{g_X}{\sqrt{1-\epsilon^2}}A'_\mu X -i g \tau^aW^a_\mu
\end{align}
after rotating the kinetic mixing term $\epsilon$ to an upper triangular gauge coupling matrix (see e.g. Ref.\ \cite{Zhang:2018fbm}). $X,Y$ and $\tau^a$ are the generators of U$(1)_X$, U$(1)_Y$ and SU$(2)_L$ respectively. The part of the Lagrangian relevant for the mass of the gauge bosons is given by
\begin{align}
    \mathcal{L}\supset \left|D_\mu H\right|^2+\left|D_\mu \zeta\right|^2.
\end{align}
Expanding this around the VEVs we obtain the neutral gauge boson mass matrix
\begin{align}
    \mathcal{L}\supset& \frac{v_H^2}{8}\left( g_Y A_\mu+\frac{-\epsilon g_Y}{\sqrt{1-\epsilon^2}}A'_\mu - g W^3_\mu\right)^2 +\frac{v_\zeta^2}{2}\left(\frac{g_X}{\sqrt{1-\epsilon^2}}A'_\mu X_\zeta\right)^2\\
    &=\left(\begin{matrix}A_\mu & W^3_\mu & A'_\mu\end{matrix}\right)\left(\begin{matrix} \frac{g_Y^2v_H^2}{8} & -\frac{gg_Yv_H^2}{8} & -\frac{\epsilon g_Y^2 v_H^2}{8\sqrt{1-\epsilon^2}} \\ -\frac{gg_Yv_H^2}{8} & \frac{g^2v_H^2}{8} & \frac{\epsilon g g_Y v_H^2}{8\sqrt{1-\epsilon^2}} \\ -\frac{\epsilon g_Y^2 v_H^2}{8\sqrt{1-\epsilon^2}}& \frac{\epsilon g g_Y v_H^2}{8\sqrt{1-\epsilon^2}} & \frac{\epsilon^2 g_Y^2 v_H^2}{8(1-\epsilon^2)}+\frac{g_X^2 v_\zeta^2 X_\zeta^2}{2(1-\epsilon^2)}\end{matrix}\right)\left(\begin{matrix}A^\mu \\ W^{3\mu} \\ A'^\mu\end{matrix}\right).
\end{align}
Diagonalizing this matrix yields one vanishing eigenvalue (for the photon) and the masses for the $Z$ and $Z'$ boson which are, up to order $\epsilon^2$, given by
\begin{align}
	m_Z^2=&M_{\tilde{Z}}^2 \left( 1 - \epsilon^2 \sin^2 \theta_w \frac{M_{\tilde{Z}}^2}{M_X^2-M_{\tilde{Z}}^2} \right), \\
	m_{Z'}^2=& M_X^2 \left(1 + \epsilon^2 \left( 1+\sin^2 \theta_w \frac{M_{\tilde{Z}}^2}{M_X^2-M_{\tilde{Z}}^2}\right)\right),
\end{align}
with the Weinberg angle $\theta_w$ and 
\begin{align}
	M_{\tilde{Z}}^2=&\frac{(g^2+g_Y^2) v_H^2}{4} =  \frac{g^2 v_H^2}{4\cos^2 \theta_w},\\
	M_X^2 =& g_X^2 X_\zeta^2 v_\zeta^2.
\end{align}

The $Z^0$ boson mass has been experimentally measured by e.g. the LEP experiments and matches the Standard Model prediction. As the shift in the $Z^0$ mass depends on the kinetic mixing and the $Z'$ mass, this can be used to constrain the viable parameter space. The mixing also modifies the couplings, allowing for further tests of the model. 

A general review of the phenomenology for heavy $Z'$ bosons can be found in Ref.\ \cite{Langacker:2008yv}. Dedicated experimental analyses on invisible $Z'$ decays have set limits in the MeV to 10 GeV range \cite{Lees:2017lec,NA64:2019imj}. Model-independent limits on the kinetic mixing have been obtained in Ref.\ \cite{Hook:2010tw}, and the phenomenology of a similar setup to ours has been discussed in \cite{Lao:2020inc}. Collider constraints have been examined in Ref.\ \cite{Aboubrahim:2022qln}. A recent review of the theory and phenomenology of massless and massive dark photons and different limits can be found in Ref.\ \cite{Fabbrichesi:2020wbt}.

\subsection{Massive neutrino generations}\label{sec:4.3}
In this section we discuss how many generations of the new fields are required in order to allow for a least two massive neutrinos. In Ref.\ \cite{Bonnet:2012kz} the formulas for the neutrino masses arising from each topology are given. These formulas are correct even in the U$(1)$ case as the fields running in the loop do not change (up to real fields becoming complex).

We first assume only one generation of each new field. In this case a simple pattern emerges in the neutrino mass formulas. Assuming that the neutrinos have Yukawa interactions with the new fields given by the couplings $y,y'$, the neutrino mass matrix is 
\begin{align}
	\left(M_\nu\right)_{\alpha_\beta}\propto y_\alpha y'_\beta+y'_\alpha y_\beta=:A_{\alpha\beta}+A_{\alpha\beta}^T.
\end{align}
The proportionality factor depends on the masses and couplings of the new fields. It is easy to see that the matrix $A$ has rank 1. The rank of $M_\nu$ and thus the number of massive neutrinos can be estimated to be $\leq2$. As the entries of the Yukawa couplings are not fixed a priori, the neutrino matrix will actually have rank two unless two fields in the loop are identified with each other, which yields $y=y'$.\footnote{Strictly speaking some entries of $y$ and $y'$ could be exactly the same or have a special relation to one another. This should not pose a problem since such a scenario would be extremely fine-tuned or would require additional symmetries.} Some models allow for the case where two fields are identified with each other. If one does so, the models predict only one massive neutrino unless one introduces several generations of one of the new fields. Identifying fields with each other is a special feature of some models which can be possible for both $Z_2$ models as well as U$(1)$ models. It is useful to observe that the $Z_2$ scotogenic model is an example where $\phi$ and $\phi'^\dagger$ are identified with each other and therefore the scotogenic model includes two (or three) generations of right handed neutrinos in order to predict two (or three) generations of massive neutrinos. Alternatively one could have $\phi$ and $\phi'$ as separate fields (or equivalently two generations of scalar doublets) and only one right handed neutrino. For the U$(1)$ version of the scotogenic model this identification is not possible.

\subsection{Dark matter}\label{sec:4.4}
The dark matter candidate can be both fermionic and scalar, whichever new (stable) field is the lightest. We do not consider $Z'$ dark matter here. Dark matter is stabililized by the residual symmetry discussed in Sec.\ \ref{sec:2.9} which is the unbroken subgroup of the U$(1)_X$ symmetry. With $\mathcal{O}(1)$ couplings and electroweak interactions, the dark sector is in thermal equilibrium in the early Universe, and the relic density is determined by the well-known freeze-out process. The dark matter self annihilation depends both on the model and the dark matter candidate and generally involves new particles in the $t$- and $u$-channels as well as neutral gauge and Higgs bosons in the $s$-channel as well as four-point interactions for scalar dark matter.

In case of neutral fermions, one might wonder whether after U$(1)$ breaking and EWSB there is a Majorana mass term. The residual symmetry also allows for some insights into this question. As it remains unbroken even after EWSB, terms forbidden by this symmetry, will not occur in the low energy theory even at loop level. If the residual symmetry is $Z_N$ with $N\neq2$, the Majorana mass term $m\psi\psi$ is forbidden by this symmetry.\footnote{In principle one might assume that, for example for a $Z_4$, the fermion transforms as $\psi\to-\psi$ as $Z_2$ is a subgroup of $Z_4$. However this case does not occur for any of the new fields in our list of models.} Conversely for a $Z_2$ symmetry one would assume that a Majorana mass term is generated at least at loop level, unless there is a further accidental symmetry present.

The case of scalar doublet dark matter requires some special attention as there is a coupling to the SM $Z$-boson $ig/(2c_W)\left(\phi^{0R}\partial^\mu\phi^{0I}-\partial^\mu\phi^{0R}\phi^{0I}\right)Z^0_\mu$, where we split the neutral component of the scalar doublet into its real and imaginary part $\phi^0=\phi^{0R}+i\phi^{0I}$. If $\phi^{0R}$ and $\phi^{0I}$ have the same mass, this interaction gives a large contribution to the direct detection cross section and thus excludes this dark matter candidate. If the residual symmetry is a $Z_N$ with $N\neq2$, then this symmetry requires a mass degeneracy of $\phi^{0R}$ and $\phi^{0I}$. If scalar dark matter is the only dark matter candidate, such a model is excluded by direct detection. For a $Z_2$ symmetry this degeneracy may be lifted at tree level or at loop level. In this case one needs to ensure that the mass splitting is larger than a couple 100 keV \cite{deBoer:2021pon}.

The dark matter particle can be searched for by standard WIMP searches such as indirect detection \cite{MAGIC:2016xys,HESS:2022ygk,IceCube:2023ies} and direct detection \cite{XENONCollaboration:2023orw,PandaX-II:2020oim,LUX-ZEPLIN:2022xrq,PICO:2019vsc}. The dark sector particles can also be searched for at colliders. The dark matter phenomenology for the $Z_2$ models has been studied in a number of models \cite{Klasen:2013jpa,deBoer:2020yyw,deBoer:2021pon,Esch:2016jyx,Esch:2018ccs,Fiaschi:2018rky,deBoer:2021xjs}. For the U$(1)$ models, the main qualitative differences are the additional annihilation and scattering channels involving the $Z'$ gauge boson and potentially the $\zeta$ Higgs boson. An analysis of direct and indirect detection dominated by kinetic mixing with a comparable setup to ours can be found in Refs.\ \cite{Mambrini:2011dw,Lao:2020inc}. It is worth noting that for $v_\zeta\sim\mathcal{O}(100\text{keV})$ the dark matter candidate can have a just slightly heavier partner, thus allowing for inelastic scattering \cite{Tucker-Smith:2001myb}.

\subsection{Further considerations}

Neutrinos with Majorana masses necessarily imply lepton number violation. In some models, the dark scalars or fermions can be assigned a specific lepton number such that only one term in the Lagrangian breaks this quantum number. This is the case in both the original and the gauged scotogenic model with the $\lambda_5$ term, when the scalars are assigned a lepton number and where this coupling is then naturally small. In the gauged model, this would be the term proportional to $(\phi'^\dagger H)(H\phi)$. In other models, this need, however, not be the case and the number of lepton number-violating terms can be larger.

Radiative seesaw models generally allow for LFV to occur through similar diagrams as the neutrino loop \cite{Toma:2013zsa,Vicente:2014wga,deBoer:2020yyw,deBoer:2021pon,Fiaschi:2018rky}. This allows for an additional way to test these models \cite{MEG:2016leq,SINDRUM:1987nra}. Similarly to the neutrino mass formula, the difference between the $Z_2$ and U$(1)$ case is limited. There may be additional Yukawa coupling in models, where one assumes fields to be identified with each other in the $Z_2$ case and in the U$(1)$ case this is not possible. Otherwise, the only difference might be additional degrees of freedom running in the loop that induces LFV processes, as real fields might become complex and Majorana fermions might become vector-like. The $Z'$-boson does not contribute to LFV at one loop.

Dark photons also contribute to the anomalous magnetic moment, given that their masses are not too heavy, and have therefore been discussed as an explanation for the muon magnetic anomaly $(g_\mu-2)$ \cite{Muong-2:2023cdq}. However this explanation is excluded by the experimental limits \cite{Lees:2017lec,NA64:2019imj} assuming invisible dark photon decays.

With new particles added, the running of the couplings is modified, which might allow for gauge coupling unification.  With the new particles being color singlets, SU$(3)_c$ running is not changed while the SU$(2)_L$ beta function becomes larger with the addition of new particles. Therefore the scale where SU$(3)_c$ and SU$(2)_L$ cross is lowered with respect to the SM only case. This has been observed in Ref.\ \cite{Hagedorn:2016dze} for the $Z_2$ case and shown for the models presented in this paper in Refs.\ \cite{deBoer:MSc,Zeinstra:PhD}. To be precise, if unification occurs, it happens at scales around $\mathcal{O}(10^{12}\text{GeV})$ to $\mathcal{O}(10^{14}\text{GeV})$. Unification scales at this order tend to be in conflict with proton decay limits \cite{Nath:2006ut,Super-Kamiokande:2020wjk}.

\section{Summary and outlook}\label{sec:5}
Minimal extensions of the Standard Model can introduce viable dark matter candidates and generate neutrino masses. To stabilize the dark sector a new symmetry must be introduced. This symmetry can be a local U$(1)$ symmetry which is theoretically better motivated than the often used discrete or global symmetries. We followed previous work and allowed for up to four dark fields in addition to the new gauge boson and the scalar Higgs fields breaking the new U$(1)$ gauge group. Arguments based on anomaly cancellation, dark matter stability and the neutrino topologies restrict the possible charge assignments. Given our assumptions, the Standard Model must be uncharged under the new gauge group and all new fermions must be vector like. All $Z_2$ models from Ref.\ \cite{Restrepo:2013aga} can be promoted to models with a gauged U$(1)$ and many models allow for different charge assignments. We listed the particle content and charge assignments of all possible models. 

The new vector boson gives rise to a rich phenomenology complementing the known phenomenology of the $Z_2$ models. The dark gauge boson, coupling to the SM simply through the kinetic mixing portal, can be massless or obtain a mass through spontaneous symmetry breaking. The kinetic mixing portal gives a model independent way of testing the modified gauge sector. In case of a massive $Z'$ boson an additional scalar field can be introduced in order to break the new U$(1)$ symmetry giving rise to signatures in the Higgs sector.

As a further direction, one could address the coincidence of scales inherent to the scotogenic models. It is assumed, that all explicit scales in these models are at 100 GeV or TeV scale. As there are a number of independent mass parameters, it is hard to explain why they all are rather close to each other. 
In a classically scale invariant setting where scale symmetry is broken dynamically \cite{Coleman:1973jx}, all scales would be related to the symmetry breaking scale. This would give an explanation for the coincidence of scales and also address the hierarchy problem \cite{Bardeen:1995kv}. Scale invariant radiative seesaw models have been presented in Refs.\ \cite{Ahriche:2016cio,Ahriche:2016ixu}. The U$(1)_X$ gauge contribution to the effective potential can easily drive dynamical symmetry breaking making our models especially attractive for a scale-invariant setting. 

\section*{Acknowledgments}

This work has been supported by the DFG through the Research Training Network 2149 ``Strong and weak interactions - from hadrons to dark matter''.

\appendix
\section{Example model T1-3-A}
As an example, we discuss the model T1-3-A with a gauged U(1) symmetry, a vector-like fermion singlet $\Psi$ and doublet $\psi$, a scalar singlet $\phi$, and hypercharge parameter $\alpha=0$. The representations and charge assignments for this concrete model are given in Tab.\ \ref{tab:T1-3-A}.

\subsection{The model}
\begin{table}
\caption{Fields contained in the simplest gauged U$(1)$ model T1-3-A.}
\label{tab:T1-3-A}
\centering
\begin{tabular}{|c|c|c|}
    \hline
    Field & Type & SU$(3)_c\times$SU$(2)_L\times$U$(1)_Y\times$U$(1)_X$\\
    \hline
    $\zeta$ & Scalar & $(1,1,0,2)$ \\
    $\phi$  & Scalar & $(1,1,0,1)$ \\
    $\Psi$ & Fermion & $(1,1,0,1)$ \\
    $\psi$& Fermion & $(1,2,-1,1)$ \\
    \hline
    $\Psi'$ & Fermion, vector partner of $\Psi$ & $(1,1,0,-1)$ \\
    $\psi'$& Fermion, vector partner of $\psi$ & $(1,2,1,-1)$ \\
    \hline
\end{tabular}
\end{table}

In addition to the terms in the SM and the terms relevant for symmetry breaking given in Sec.\ \ref{sec:H_sector}, we have the following terms in the Lagrangian:
\begin{align}
    \mathcal{L}_\text{Ferm}\supset& - M_{\psi\psi'} \psi \psi' - M_{\Psi\Psi'} \Psi\Psi' 
 - y_{1} \left({H}^\dagger \psi'\right) \Psi - y_{2} {\zeta}^\dagger \Psi \Psi  - \left(y_{3} H \psi\right) \Psi' - y_{4} \zeta \Psi' \Psi' \nonumber \\
 &- y_{5} \left(L \psi'\right) \phi + \text{h.c.},  \\
 V\supset&  M_{\phi}^{2} |\phi|^2
 - \left(\frac{\kappa}{2} {\zeta}^\dagger {\phi}^2 + \text{h.c.}\right)
 - \lambda_{1} |H|^2|\phi|^2 - \frac{\lambda_{2}}{2} |\phi|^4 - \lambda_{3} |\zeta|^2|\phi|^2.
\end{align}
After symmetry breaking, the dark scalar is split into two real components $\phi=\frac{1}{\sqrt{2}}(\phi_R+i\phi_I)$ with masses
\begin{align}
    m_{\phi_R}^2=&M_\phi^2+\lambda_1\frac{v_H^2}{2}+\lambda_3\frac{v_\zeta^2}{2}+\kappa\frac{v_\zeta}{\sqrt{2}},\\
    m_{\phi_I}^2=&M_\phi^2+\lambda_1\frac{v_H^2}{2}+\lambda_3\frac{v_\zeta^2}{2}-\kappa\frac{v_\zeta}{\sqrt{2}},
\end{align}
where we assumed $\kappa$ to be real. In the case where $\kappa v_\zeta\sim {\cal O}(100\text{ keV})^2$, it is possible for inelastic scattering off nucleons to occur via a $Z'$-boson exchange. The neutral dark fermions in the basis $(\Psi,\Psi',\psi,\psi')$ have the following mass matrix after symmetry breaking:
\begin{align}
    M_0=\left(\begin{matrix}
    \sqrt{2}y_2v_\zeta & M_{\Psi\Psi'} & 0 & y_1\frac{v}{\sqrt{2}}\\ 
    M_{\Psi\Psi'} & \sqrt{2}y_4v_\zeta & -y_3\frac{v}{\sqrt{2}} &0 \\
    0 & -y_3\frac{v}{\sqrt{2}} & 0 & M_{\psi\psi'} \\
    y_1\frac{v}{\sqrt{2}} & 0 & M_{\psi\psi'} &0     
    \end{matrix}\right).
\end{align}
The Yukawa couplings $y_{2,4}$ turn the neutral fermions into Majorana fermions. We assume this matrix is diagonalized by a unitary matrix $U_\chi$ as $U^*_\chi M_0U^\dagger_\chi=\text{diag}(m_{\chi_1},m_{\chi_2},m_{\chi_3},m_{\chi_4})$.

\subsection{Neutrino mass}
\begin{figure}
    \centering
    \includegraphics{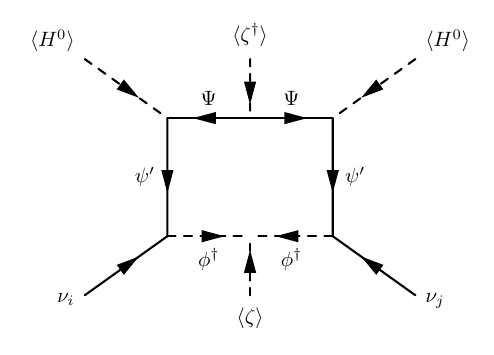}
    \caption{Lowest-order neutrino mass diagram in model T1-3-A ($\alpha=0$) with explicit $\zeta$ insertions.}
    \label{fig:T1-3-A_zeta}
\end{figure}

\begin{figure}
    \centering
    \includegraphics{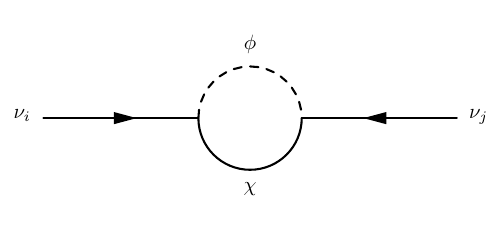}
    \caption{Neutrino mass diagram after EWSB.}
    \label{fig:self_energy}
\end{figure}

The formula for the neutrino masses are sometimes calculated keeping only the lowest orders of interactions with the VEVs (see e.g. Ref.\ \cite{Bonnet:2012kz}). For the T1-3 topology, there are always at least four propagators, and the loop integral is therefore explicitly finite. The lowest-order neutrino mass diagram for the model discussed in this section is shown in Fig.\ \ref{fig:T1-3-A_zeta}. 
However, one can also resum all VEV insertions by treating them as parts of the mass matrix. The corresponding diagram is shown in Fig.\ \ref{fig:self_energy}. 
The diagram with vanishing external momenta can be evaluated to 
\begin{align}
    i(M_\nu)^{\alpha\beta}=\sum_i \frac{y_5^\alpha (U_\chi^*)^{i4}y_5^\beta (U_\chi^*)^{i4}}{2}m_{\chi_i}\int\frac{d^4k}{(2\pi)^4}\frac{1}{(k^2-m_{\chi_i}^2)(k^2-m_{\phi_R}^2)}-\left(R\to I\right).
\end{align}
The loop integral is then
\begin{align}\label{eq:A.7}
    \int\frac{d^4k}{(2\pi)^4}&\frac{1}{(k^2-m_{\chi_i}^2)(k^2-m_{\phi_R}^2)}= \\
    &=\frac{i}{16 \pi^2}\left\{\Delta+1-\frac{1}{m_{\chi_i}^2-m_{\phi_R}^2}\left[m_{\chi_i}^2\ln\left(\frac{m_{\chi_i}^2}{\mu^2}\right)-m_{\phi_R}^2\ln\left(\frac{m_{\phi_R}^2}{\mu^2}\right)\right]\right\}\nonumber 
\end{align}
with $\Delta=\frac{1}{\epsilon}-\gamma_E+\ln 4\pi$. The neutrino mass matrix is then 
\begin{align}
    (M_\nu)^{\alpha\beta}=\sum_i \frac{y_5^\alpha (U_\chi^*)^{i4}y_5^\beta (U_\chi^*)^{i4}}{32\pi^2}\frac{m_{\chi_i}}{m_{\phi_R}^2\!-\!m_{\chi_i}^2}\left[m_{\chi_i}^2\!\ln\left(\!\frac{m_{\chi_i}^2}{\mu^2}\!\right)\!-\!m_{\phi_R}^2\!\ln\left(\!\frac{m_{\phi_R}^2}{\mu^2}\!\right)\right]\!-\!\left(R\to I\right).
\end{align}
As discussed in Sec.\ \ref{sec:4.3}, this matrix has only rank one, and therefore there is only one massive neutrino. In order two obtain at least two generations of massive neutrinos, one can either introduce two generations of $\psi$ or two generations of $\phi$. The generalization of the neutrino mass formula is straightforward. E.g., for two generations of scalar fields and assuming diagonal couplings for simplicity one obtains
\begin{align}
    (M_\nu)^{\alpha_\beta}\!=\!\sum_{ij} \frac{y_5^{\alpha j} (U_\chi^*)^{i4}y_5^{\beta j} (U_\chi^*)^{i4}}{32\pi^2}\frac{m_{\chi_i}}{m_{\phi_{jR}}^2\!-\!m_{\chi_i}^2}\left[m_{\chi_i}^2\!\ln\left(\!\frac{m_{\chi_i}^2}{\mu^2}\!\right)\!-\!m_{\phi_{jR}}^2\!\ln\left(\!\frac{m_{\phi_{jR}}^2}{\mu^2}\!\right)\right]\!-\!\left(R\to I\right).
\end{align}
It is interesting to notice that if any of the sets of couplings $y_{1,3}$, $y_{2,4}$, $y_5$ or $\kappa$ are zero, then the neutrino masses vanish, since at least one of the vertices in Fig.\ \ref{fig:T1-3-A_zeta} is forbidden.

\section{T3-B}
In this section, we briefly present the U$(1)$ version of the scotogenic model, i.e.\ model T3-B with a vector-like fermion singlet $\psi$, two scalar doublets $\phi,\phi'$ and $\alpha=-1$. This model has been proposed in Ref.\ \cite{Ma:2013yga}. However, the full Lagrangian is not given in this reference.

\subsection{The model}

\begin{table}
\caption{Fields contained in the gauged U$(1)$ model T3-B}
\label{tab:T3-B}
\centering
\begin{tabular}{|c|c|c|}
    \hline
    Field & Type & SU$(3)_c\times$SU$(2)_L\times$U$(1)_Y\times$U$(1)_X$\\
    \hline
    $\zeta$ & Scalar & $(1,1,0,2)$ \\
    $\phi$  & Scalar & $(1,2,1,\beta)$ \\
    $\phi'$ & Scalar & $(1,2,-1,\beta)$ \\
    $\psi$ & Fermion & $(1,1,0,\beta)$ \\
    \hline
    $\psi'$ & Fermion, vector partner of $\psi$ & $(1,1,0,-\beta)$ \\
    \hline
\end{tabular}
\end{table}
The Lagrangian involving the new fermions is given by
\begin{align}
    \mathcal{L}\supset -M_{\psi\psi'}\psi\psi'-y_1\zeta^\dagger\psi\psi-y_2(\phi^\dagger L)\psi-y_3\zeta\psi'\psi'-y_4 (L\phi')\psi'+\text{h.c.}
\end{align}
The scalar potential up to the terms discussed in Sec.\ \ref{sec:H_sector} is given by
\begin{align}
    V\supset& M_{\phi}^{2} {\phi}^\dagger \phi + M_{\phi'}^{2} {\phi'}^\dagger \phi' + \lambda_{1} \left({H}^\dagger H\right) \left({\phi}^\dagger \phi\right) + \lambda_{2} \left({H}^\dagger H\right) \left({\phi'}^\dagger \phi'\right) \nonumber\\
&+ \lambda_{3} \left({H}^\dagger \phi'\right) \left({\phi'}^\dagger H\right) + \lambda_{4} \left({H}^\dagger \phi\right) \left({\phi}^\dagger H\right) + \lambda_{5} \left({\phi}^\dagger \phi'\right) \left({\phi'}^\dagger \phi\right) + \frac{\lambda_{6}}{2} \left({\phi}^\dagger \phi\right)^2\\
&+ \lambda_{7} \left({\phi}^\dagger \phi\right) \left({\phi'}^\dagger \phi'\right) + \frac{\lambda_{8}}{2} \left({\phi'}^\dagger \phi'\right)^2 + \lambda_{9} \left({\phi'}^\dagger \phi'\right) {\zeta}^\dagger \zeta  + \lambda_{10} \left({\phi}^\dagger \phi\right) {\zeta}^\dagger \zeta \nonumber\\
&+ \left(\lambda_{11} \left({\phi'}^\dagger H\right) \left(H \phi\right) + \text{H.\,c.}\right)+ \left(\kappa \left(\phi \phi'\right) {\zeta}^\dagger + \text{h.c.}\right).\nonumber
\end{align}
Note that the couplings $y_{1,3}$ and $\kappa$ are only possible for $\beta=1$. Neutrino masses are also generated in the case of $\beta\neq1$.

Assuming real $\kappa$ and $\lambda_{11}$, the neutral scalar mass matrix $M_{\phi^0}^2$ in the basis $(\phi_R,\phi_I,\phi'_R,\phi'_I)$ is given by 
\begin{align}
    \left(\begin{matrix}
        M_\phi^2+\lambda_1\frac{v_H^2}{2}+\lambda_{10}\frac{v_\zeta^2}{2} & 0 & -\lambda_{11}\frac{v_H^2}{2}+\kappa\frac{v_\zeta}{\sqrt{2}} & 0 \\
        0 & M_\phi^2+\lambda_1\frac{v_H^2}{2}+\lambda_{10}\frac{v_\zeta^2}{2} & 0 &-\lambda_{11}\frac{v_H^2}{2}-\kappa\frac{v_\zeta}{\sqrt{2}}\\
        -\lambda_{11}\frac{v_H^2}{2}+\kappa\frac{v_\zeta}{\sqrt{2}} & 0 & M_{\phi'}^2+(\lambda_2+\lambda_3)\frac{v_H^2}{2}+\lambda_{9}\frac{v_\zeta^2}{2} & 0 \\
        0 & -\lambda_{11}\frac{v_H^2}{2}-\kappa\frac{v_\zeta}{\sqrt{2}} & 0 & M_{\phi'}^2+(\lambda_2+\lambda_3)\frac{v_H^2}{2}+\lambda_{9}\frac{v_\zeta^2}{2}
    \end{matrix}\right).
\end{align}
Note that for vanishing $\kappa$, the neutral scalar fields remain complex. This matrix is diagonalized by an orthogonal matrix $O_\phi$ as $O_\phi M_{\phi^0}^2O_\phi^T=\text{diag}(m_{\eta_1}^2,m_{\eta_2}^2,m_{\eta_3}^2,m_{\eta_4}^2)$. The mass matrix for the charged scalar fields is given by
\begin{align}
    \left(\begin{matrix}
        M_\phi^2+(\lambda_1+\lambda_4)\frac{v_H^2}{2}+\lambda_{10}\frac{v_\zeta^2}{2} & -\kappa\frac{v_\zeta}{\sqrt{2}} \\
        -\kappa\frac{v_\zeta}{\sqrt{2}} & M_{\phi'}^2+\lambda_2\frac{v_H^2}{2}+\lambda_{9}\frac{v_\zeta^2}{2} 
    \end{matrix}\right).
\end{align}
The (neutral) fermion mass matrix in the basis $(\psi,\psi')$ is given by
\begin{align}
   M_0= \left(\begin{matrix}
        \sqrt{2} y_1 v_\zeta & M_{\psi\psi'} \\
        M_{\psi\psi'} & \sqrt{2} y_3 v_\zeta
    \end{matrix}\right).
\end{align}
For vanishing $y_{1,3}$, the fermion does not turn into a Majorana fermion, but remains vector-like. Again we diagonalize $M_0$ with a unitary matrix $U_\chi$ as $U^*_\chi M_0U^\dagger_\chi=\text{diag}(m_{\chi_1},m_{\chi_2})$.

\subsection{Neutrino mass}

The neutrino mass can again be computed from the diagram in Fig.\ \ref{fig:self_energy}. We find
\begin{align}
    i(M_\nu)^{\alpha\beta}=&\sum_{i,j} \frac{1}{2}\left(y_2^\alpha(U_\chi^*)^{i1}\left(O_\phi^{j1}-iO_\phi^{j2}\right)+y_4^\alpha(U_\chi^*)^{i2}\left(O_\phi^{j4}+iO_\phi^{j4}\right)\right)\times\nonumber\\
    &\qquad\times\left(y_2^\beta(U_\chi^*)^{i1}\left(O_\phi^{j1}-iO_\phi^{j2}\right)+y_4^\beta(U_\chi^*)^{i2}\left(O_\phi^{j4}+iO_\phi^{j4}\right)\right)\times\\
    &\qquad\times m_{\chi_i}\int\frac{d^4k}{(2\pi)^4}\frac{1}{(k^2-m_{\chi_i}^2)(k^2-m_{\eta_j}^2)}\nonumber.
\end{align}
Using Eq.\ (\ref{eq:A.7}) to evaluate the loop integral, we find
\begin{align}
    (M_\nu)^{\alpha\beta}=&\sum_{ij} \frac{1}{32\pi^2}\left(y_2^\alpha(U_\chi^*)^{i1}\left(O_\phi^{j1}-iO_\phi^{j2}\right)+y_4^\alpha(U_\chi^*)^{i2}\left(O_\phi^{j4}+iO_\phi^{j4}\right)\right)\times\nonumber\\
    &\qquad\times\left(y_2^\beta(U_\chi^*)^{i1}\left(O_\phi^{j1}-iO_\phi^{j2}\right)+y_4^\beta(U_\chi^*)^{i2}\left(O_\phi^{j4}+iO_\phi^{j4}\right)\right)\times\\
    &\qquad\times\frac{m_{\chi_i}}{m_{\eta_j}^2-m_{\chi_i}^2}\left[m_{\chi_i}^2\ln\left(\frac{m_{\chi_i}^2}{\mu^2}\right)-m_{\eta_j}^2\ln\left(\frac{m_{\eta_j}^2}{\mu^2}\right)\right].\nonumber
\end{align}
The constant (non-logarithmic) and divergent terms cancel as $O_\phi O_\phi^T=I$.

{\footnotesize{}\bibliographystyle{JHEP.bst}
\bibliography{bib.bib}
}{\footnotesize\par}
\end{document}